\documentstyle[11pt]{article}

\title{Modal-type orthomodular logic}

\author{{\sc G. Domenech\thanks{Fellow of the Consejo Nacional de
Investigaciones Cient\'ificas y T\'ecnicas} $^{,1,3}$ , H.
Freytes$^{*,2}$ and C. de Ronde$^{3,4}$}}

\date{{\small 1. Instituto de Astronom\'ia  y F\'isica del Espacio (IAFE)\\
Casilla de Correo 67, Sucursal 28, 1428 Buenos Aires, Argentina \\
2. Instituto Argentino de Matem\'atica (IAM)\\
Saavedra 15 - 3er Piso - 1083  Buenos Aires, Argentina \\
3. Center Leo Apostol (CLEA)\\
4. Foundations of the Exact Sciences (FUND)\\
Brussels Free University - Krijgskudestraat 33, 1160
Brussels-Belgium }}

\begin{document}

\bibliographystyle{plain}

\maketitle

\begin{abstract}

\noindent In this paper we enrich the orthomodular structure by
adding a modal operator, following a physical motivation. A logical
system is developed, obtaining algebraic completeness and
completeness with respect to a Kripke-style semantic founded on Baer
$^\star$-semigroups as in \cite{YM}.

\end{abstract}

\begin{small}



\end{small}

\bibliography{pom}

\begin{thebibliography}{10}

\bibitem{BV} G. Birkhoff, and J. von Neumann,  {\it The logic of quantum mechanics}, Ann. Math. {\bf 27}  (1936) 823-843.

\bibitem{BH} G. Bruns and J. Harding, {\it Algebraic aspects of orthomodular lattices}, in ``Current Research in Operational Quantum Logic'',
B. Coecke, D. Moore and A. Wilce (Eds.), Kluwer Academic Publishers,
Dordrecht, 2000.

\bibitem{Bur} S. Burris and  H. P.  Sankappanavar, ``A Course in Universal
Algebra'', Graduate Text in Mathematics,  Springer-Verlag, New York,
1981.

\bibitem{D88} D. Dieks, {\it The formalism of quantum theory: an
objective description of reality}, Ann. der Physik {\bf 7} (1988)
174-190.

\bibitem{D89} D. Dieks, {\it Quantum mechanics without the
projection postulate and its realistic interpretation}, Found. Phys.
{\bf 19} (1989) 1397-1423.

\bibitem{D05} D. Dieks, {\it Quantum mechanics: an intelligible description of objective
reality?},  Found.  Phys. {\bf 35}  (2005) 399-415.

\bibitem{DIS}  H. Dishkant, {\it Imbedding of the quantum logic in the modal systems of Brower}, J. Symbolic Logic {\bf 42} (1977) 421-328.

\bibitem{DF} G. Domenech and H. Freytes, {\it  Contextual logic for quantum
systems}, J. Math. Phys. {\bf 46} (2005) 012102-1 - 012102-9.

\bibitem{DFD1} G. Domenech, H. Freytes and C. de Ronde,  {\it Scopes and limits of modality in quantum
mechanics}, Ann. der Physik, {\bf 15} (2006) 853-860.

\bibitem{DFD2} G. Domenech, H. Freytes and C. de Ronde, {\it A topological study of contextuality and modality in
quantum mechanics}, Int. J. Theor. Phys., (2007) in press.


\bibitem{FOU} D. J. Foulis,  {\it Baer $^{\star}$-semigroups}, Proccedings of American Mathematical Society {\bf 11}  (1960)
648-654.

\bibitem{GOLD} R. I. Goldblatt,  {\it Orthomodularity is not elemntary}, J. Symbolic Logic {\bf 49} (1984) 401-404.

\bibitem{HAR}  G. Hardegree, {\it Material implication in orthomodular (and Boolean) lattices}, Notre Dame Journal of Formal Logic {\bf 22}  (1981)
163-181.

\bibitem{HP}  L. Herman and R. Piziak,  {\it Modal propositional logic on an otrhomodular basis I}, J. Symbolic Logic {\bf 39} (1974)
478-488.

\bibitem{JAN}  M. F. Janowitz, {\it Quantifiers and orthomodular lattices} Pacific J. Math {\bf 13}  (1963)
660-676.

\bibitem{Ka} J. A. Kalman, {\it Lattices with involution}, Trans. Amer. Math. Soc.  87 (1958) 485-491.

\bibitem{KAL}  G. Kalmbach,  ``Ortomodular Lattices'', Academic Press, London, 1983.

\bibitem{KS}   S. Kochen  and  E. P. Specker,  {\it The problem of hidden variables
in quantum mechanics}, J. Math. Mech. {\bf 17} (1967) 9-87.

\bibitem{MM}  F. Maeda, S. Maeda, ``Theory of Symmetric Lattices'', Springer-Verlag,  Berlin, 1970.

\bibitem{YM} Y. Miyazaki,  {\it Kripke-style semantics of orthomodular logics}, Math. Log. Quart. {\bf 47} (2001)  341-362.

\bibitem{VIO} V. Sofronie-Stokkermans, {\it Representation theorems and the semantics of
non-classical logics, and applications to automated theorem
proving}, in ``Theory and Applications of Multiple-Valued Logic'',
M. Fitting and E. Orlowska (Eds.), Springer-Verlag Series Studies in
Fuzziness and Soft Computing, 2003.

\bibitem{vF73} B. C. van Fraassen, {\it Semantic analysis of quantum logic},
in ``Contemporary Research in the Foundations and Philosophy of
Quantum Theory'', C. A. Hooker (Ed.), Reidel, Dordrecht, 1973.



\end{thebibliography}

\newtheorem{theo}{Theorem}[section]

\newtheorem{definition}[theo]{Definition}

\newtheorem{lem}[theo]{Lemma}

\newtheorem{prop}[theo]{Proposition}

\newtheorem{coro}[theo]{Corollary}

\newtheorem{exam}[theo]{Example}

\newtheorem{rema}[theo]{Remark}{\hspace*{4mm}}

\newtheorem{example}[theo]{Example}

\newcommand{\proof}{\noindent {\em Proof:\/}{\hspace*{4mm}}}

\newcommand{\qed}{\hfill$\Box$}

\section*{Introduction}

In their 1936 seminal paper \cite{BV}, Birkhoff and von Neumann made
the proposal of a non-classical logic for quantum mechanics founded
on the basic lattice-order properties of all closed subspaces of a
Hilbert space. This lattice-order properties are captured in the
orthomodular lattice structure. The orthomodular structure is
characterized by a weak form of distributivity called {\it
orthomodular law}. This ``weak distributivity'', which is the
essential difference with the Boolean structure, makes it extremely
intractable in certain aspects. In fact, a general representation
theorem for a class of algebras, which has as particular instances
the representation theorems as algebras of sets for Boolean algebras
and distributive lattices, allows in many cases and in a uniform way
the choice of a Kripke-style model and to establish a direct
relationship with the algebraic model \cite{VIO}. In this procedure
the distributive law plays a very important role. In absence of
distributivity this general technique is not applicable,
consequently to obtain  Kripke-style semantics may be complicated.
Such is the case for the orthomodular logic. Indeed, in \cite{GOLD},
Goldblatt gives a Kripke-style semantic for the orthomodular logic
based on an imposed restriction  on the Kripke-style semantic for
the orthologic. This restriction is not first order expressible.
Thus the obtained semantic is not very attractive. In \cite{YM},
Miyazaki introduced another approach to the Kripke-style semantic
for the orthomodular logic based on the representation theorem by
Baer semigroups given by Foulis in \cite{FOU} for orthomodular
lattices. In this way a Kripke-style model is obtained whose
universe is given by semigroups with additional operations.

Several authors added modal enrichments to the orthomodular
structure based on generalizations of classic modal systems
\cite{DIS,HAR,HP}, or generalization of quantifiers in the sense of
Halmos \cite{JAN}. In \cite{DFD1} and \cite{DFD2}, we have
introduced an orthomodular structure enriched with a modal operator
called {\it Boolean saturated orthomodular lattice}. This structure
has a rigorous physical motivation and allows to establish
algebraic-type versions of the Born rule and the well known
Kochen-Specker (KS) theorem \cite{KS}.

The aim of this paper is to study this structure from a
logic-algebraic perspective. The paper is structured as follows.
Section 1 contains generalities on universal algebra, orthomodular
lattices and Baer $^\star$-semigroups. In section 2, the physical
motivation for the modal enrichment of the orthomodular structure is
presented. In section 3 we introduce the class of Boolean saturated
orthomodular lattices ${\cal OML}^\Box$ and we prove that this class
conforms a discriminator variety. In section 4, a Hilbert-style
calculus is introduced obtaining a strong completeness theorem for
the variety ${\cal OML}^\Box$. Finally, in section 5, we give a
representation theorem by means of a sub-class of Baer
$^\star$-semigroups for ${\cal OML}^\Box$. This allows to develop a
Kripke-style semantic for the calculus of the precedent section
following the approach given in \cite{YM}. A strong completeness
theorem for these Kripke-style models  is also obtained.

\section{Basic notions}
We freely use all basic notions of universal algebra that can be
found in \cite{Bur}. If $K$ is a class of algebras of the same type
then we denote by ${\cal V}(K)$ the variety generated by $K$. Let
${\cal A}$ be a variety of algebras of type $\sigma$. We denote by
$Term_{\cal A}$ the {\it absolutely free algebra} of type $\sigma$
built  from the set of variables $V = \{x_1, x_2,...\}$. Each
element of $Term_{\cal A}$ is referred to as a {\it term}. We denote
by $Comp(t)$ the complexity of the term $t$. Let $A \in {\cal A}$. If $t \in
Term_{\cal A}$ and $a_1,\dots, a_n \in A$, by $t^A(a_1,\dots, a_n)$
we denote the result of the application of the term operation $t^A$
to the elements $a_1,\dots, a_n$. A {\it valuation} in $A$ is a map
$v:V\rightarrow A$.
Of course, any valuation $v$ in $A$ can be uniquely extended to an
${\cal A}$-homomorphism $v:Term_{\cal A} \rightarrow A$ in the usual way, i.e.,
if $t_1, \ldots, t_n \in Term_{\cal A}$ then $v(t(t_1, \ldots, t_n)) = t^A(v(t_1), \ldots, v(t_n))$. Thus, valuations are identified
with ${\cal A}$-homomorphisms from the absolutely free algebra. If $t,s \in Term_{\cal A}$, $\models_A t = s$ means
that for each valuation $v$ in $A$, $v(t) = v(s)$ and $\models_{{\cal A}} t=s$ means that for each
$A\in {\cal A}$, $\models_{A} t = s$. We denote by $Con(A)$ the lattice of congruences
of $A$. A {\it discriminator term} for $A$ is a term $t(x,y,z)$ such that $$ t^A(x,y,z) = \cases {x, & if $x\not=y$
\cr z , & if $x=y$\cr} $$ The variety ${\cal A}$ is a {\it discriminator variety} iff there exists
a class of algebras $K$ with a common discriminator term $t(x,y,z)$ such that ${\cal A} = {\cal V}(K)$. \\

Now we recall from \cite{KAL} and \cite{MM} some notions about
orthomodular lattices. A {\it  lattice with involution} \cite{Ka} is
an algebra $\langle L, \lor, \land, \neg \rangle$ such that $\langle
L, \lor, \land \rangle$ is a  lattice and $\neg$ is a unary
operation on $L$ that fulfills the following conditions: $\neg \neg
x = x$ and $\neg (x \lor y) = \neg x \land \neg y$. Let $L=\langle
L,\lor,\land, 0, 1\rangle$ be a bounded lattice. Given $a, b, c$ in
$L$, we write: $(a,b,c)D$\ \   iff $(a\lor b)\land c = (a\land
c)\lor (b\land c)$; $(a,b,c)D^{*}$ iff $(a\land b)\lor c = (a\lor
c)\land (b\lor c)$ and $(a,b,c)T$\ \ iff $(a,b,c)D$, (a,b,c)$D^{*}$
hold for all permutations of $a, b, c$. An element $z$ of a lattice
$L$ is called {\it central} iff for all elements $a,b\in L$ we have\
$(a,b,z)T$. We denote by $Z(L)$ the set of all central elements of
$L$ and it is called the {\it center} of $L$. $Z(L)$  is a Boolean
sublattice of $L$ {\rm \cite[Theorem 4.15]{MM}}. An {\it
orthomodular lattice} is an algebra $\langle L, \land, \lor, \neg,
0,1 \rangle$ of type $\langle 2,2,1,0,0 \rangle$ that satisfies the
following conditions

\begin{enumerate}
\item
$\langle L, \land, \lor, \neg, 0,1 \rangle$ is a bounded lattice with involution,

\item
$x\land  \neg x = 0 $.

\item
$x\lor ( \neg x \land (x\lor y)) = x\lor y $

\end{enumerate}

We denote by ${\cal OML}$ the variety of orthomodular lattices. An important characterization of the equations in ${\cal OML}$ is given by:

\begin{equation}\label{ECMOL}
\models_{{\cal MOL}} t = s \hspace{0.4cm} iff  \hspace{0.4cm} \models_{{\cal
MOL}} (t\land s) \lor (\neg t \land  \neg s ) = 1
\end{equation}

Therefore we can safely assume that all ${\cal OML}$-equations are of the form $t = 1$, where $t \in Term_{\cal OML}$.

\begin{rema}
{\rm  It is clear that this characterization is maintained for each
variety ${\cal A}$ such that there are terms of the language of
${\cal A}$ defining on each $A\in {\cal A}$ operations $\lor$,
$\land$, $\neg$, $0,1$ such that $L(A)=\langle A,\lor,\land, \neg,
0,1\rangle$ is an orthomodular lattice. }
\end{rema}

\begin{prop}\label{eqcentro} {\rm \cite[Lemma 29.9 and Lemma 29.16]{MM}}
Let $L$ be an orthomodular lattice then we have that:

\begin{enumerate}
\item
$z \in Z(L)$ if and only if $a = (a\land z) \lor (a \land \neg z)$ for each $a\in L$.

\item
If $L$ is complete then, $Z(L)$ is a complete lattice and for each
family $(z_i)$ in $Z(L)$ and $a \in L$, $a\land \bigvee z_i =
\bigvee (a\land z_i)$. \qed

\end{enumerate}

\end{prop}

Now we recall from \cite{FOU}, \cite{MM} and \cite{YM}  some notions
about Baer $^\star$-semigroups. A {\it $^\star$-semigroup} is an
algebra $\langle G, \cdot , \star , 0 \rangle$ of type $\langle
2,1,0 \rangle$ that satisfies the following equations:

\begin{enumerate}
\item
$\langle G, \cdot \rangle$ is a semigroup

\item
$0\cdot x = x \cdot 0 = 0$,

\item
$(x \cdot y)^\star = y^\star \cdot x^\star$,

\item
$x^{\star \star} = x $.
\end{enumerate}

Let $G$ be a $^\star$-semigroup. An element $e\in G$ is a {\it
projection} iff $e = e^\star = e\cdot e$. The set of all projections
of $G$ is denoted by $P(G)$.  Let $M$ be a non empty subset of $G$.
If $x\in G$ we define $x \cdot M = \{x \cdot m \in G: m \in M \} $
and $M \cdot x = \{m \cdot x \in G: m \in M \} $. Moreover $x$ is
said to be a {\it left annihilator} of $M$ iff $x\cdot M = \{ 0 \}$
and it is said to be a {\it right annihilator} of $M$ iff $M\cdot x
= \{ 0 \}$. We denote by $M^l$ the set of left annihilators of $M$
and by $M^r$ the set of right annihilators of $M$. A
$^\star$-semigroup is called {\it Baer $^\star$-semigroup} iff for
each $x\in G$ there exists $e\in G$ such that $$\{x\}^r = \{y \in G
: x\cdot y = 0 \} = e \cdot G$$

We do not assume in general that any $e\in P(G)$ can be represented
as $\{x\}^r = e \cdot G$ for some $x\in G$. Thus we say that $e\in
P(G)$ is a {\it closed projection} iff there exists $x\in G$ such
that $\{x\}^r = e\cdot G$. The set of all closed projections is
denoted by $P_c(G)$. Let $G$ be an orthomodular frame. From  {\rm
\cite[Lemma 37.2]{MM}}, for each $x\in G$ there exists a unique
projection $e_x\in P(G)$ such that $\{x\}^r = e \cdot G$. We denote
this $e_x$ by $x'$.  Moreover $0'$ is denoted as $1$. We can define
a partial order $\langle P(G), \leq \rangle$ as follows: $$e \leq f
\Longleftrightarrow e \cdot f = e$$

\begin{prop}\label{PROPG}
Let $G$ be a Baer $^\star$-semigroup. For any $e_1, e_2 \in P_c(G)$,
we have that:

\begin{enumerate}
\item
$e \leq f$ \hspace{0.2cm} iff \hspace{0.2cm} $e\cdot G \subseteq f\cdot G$ \hspace{0.2cm} {\rm \cite[Theorem 37.2]{MM}}

\item
$x\cdot 1 = 1 \cdot x = x$ \hspace{1.9cm} {\rm \cite[Theorem
37.4]{MM}} \qed
\end{enumerate}

\end{prop}

\begin{theo}\label{PRO1}{\rm \cite[Theorem 37.8]{MM}}
Let $G$ be a Baer $^\star$-semigroup. For any $e_1, e_2 \in P_c(G)$,
we define the following operation:

\begin{enumerate}
\item
$e_1 \land e_2 = e_1\cdot (e_2' \cdot e_1)'$,

\item
$e_1 \lor e_2 = (e_1' \land e_2')'$.

\end{enumerate}

\noindent
then $\langle P_c(G), \land, \lor, ', 0,1  \rangle$ is an orthomodular lattice with respect to the order $\langle P(G), \leq \rangle$.
\qed
\end{theo}

On the other hand we can build a Baer $^\star$-semigroup from a
partial ordered set. Let $\langle A, \leq  \rangle$ be a partial
ordered set. If $\varphi: A \rightarrow A$ is an order homomorphism
then, a {\it residual map} for $\varphi$ is an order homomorphism
$\varphi^\natural: A \rightarrow A$ such that $(\varphi \circ
\varphi^\natural)(x) \leq x \leq (\varphi^\natural \circ
\varphi)(x)$ where $\circ$ is the composition of
order-homomorphisms. We denote by $G(A)$ the set of
order-homomorphisms in $A$ admitting residual maps. If we consider
the constant order-homomorphism $\theta$, given by $\theta(x) = 0$,
then $\theta \in G(A)$ and $\langle G(A), \circ, \theta  \rangle$ is
a semigroup.

\begin{theo}\label{PRO2}{\rm \cite[Theorem 8]{FOU}}
Let $A$ be an orthomodular lattice and we consider the semigroup
$\langle G(A), \circ, \theta  \rangle$. If for each $\varphi \in
G(A)$ we define $\varphi^\star$ as $\varphi^\star(x) = \neg
\varphi^\natural(\neg x)$ then we have that:

\begin{enumerate}
\item
$\langle G(A), \circ, \star, \theta \rangle$ is a Baer
$^\star$-semigroup,

\item
if we define $\mu_a(x) = (x\lor \neg a) \land a$ for each $a\in A$ then $P_c(G(A)) = \{\mu_a : a \in A \}$,

\item
$f:A \rightarrow P_c(G(A))$ such that $f(a) = \mu_a$ is a ${\cal
OML}$-homomorphism. \qed

\end{enumerate}
\end{theo}

\section{Physical motivation of the modally enriched orthomodular structure}

In the usual terms of quantum logic, a property of a system is
related to a subspace of the Hilbert space ${\mathcal H}$ of its
(pure) states or, analogously, to the projector operator onto that
subspace. A physical magnitude ${\mathcal M}$ is represented by an
operator $\bf M$ acting over the state space. For bounded
self-adjoint operators, conditions for the existence of the spectral
decomposition ${\bf M}=\sum_{i} a_i {\bf P}_i=\sum_{i} a_i
|a_i><a_i|$ are satisfied. The real numbers $a_i$ are related to the
outcomes of measurements of the magnitude ${\mathcal M}$ and
projectors $|a_i><a_i|$ to the mentioned properties.  Thus, the
physical properties of the system are organized in the lattice of
closed subspaces ${\mathcal L}({\mathcal H})$. Moreover, each
self-adjoint operator $\bf M$  has associated a Boolean sublattice
$W_M$ of $L({\mathcal H})$ which we will refer to as the spectral
algebra of the operator $\bf M$.

Assigning values to a physical quantity ${\cal M}$ is equivalent to
establishing a Boolean homomorphism $v: W_M \rightarrow {\bf 2}$,
being ${\bf 2}$ the two elements Boolean algebra. Thus we can say
that it makes sense to use the ``classical discourse'' --this is,
the classical logical laws are valid-- within the context given by
${\mathcal M}$.

One may define a {\it global valuation} of the physical magnitudes
over ${\mathcal L}({\mathcal H})$ as a family of Boolean
homomorphisms $(v_i: W_i \rightarrow {\bf 2})_{i\in I}$ such that
$v_i\mid W_i \cap W_j = v_j\mid W_i \cap W_j$ for each $i,j \in I$,
being $(W_i)_{i\in I}$ the family of Boolean sublattices of
${\mathcal L}({\mathcal H})$. This global valuation would give the
values of all magnitudes at the same time maintaining a {\it
compatibility condition} in the sense that whenever two magnitudes
shear one or more projectors, the values assigned to those
projectors are the same from every context. As we have proved in
\cite{DF}, the KS theorem in the algebraic terms of the previous
definition rules out this possibility:

\begin{theo}\label{CS3}
If $\mathcal{H}$ is a Hilbert space such that $dim({\cal H}) > 2$,
then a global valuation over ${\mathcal L}({\mathcal H})$ is not
possible.\qed
\end{theo}

This impossibility to assign values to the properties at the same
time satisfying compatibility conditions is a weighty obstacle for
the interpretation of the formalism. B. van Fraassen was the first
one to formally include the reasoning of modal logic to circumvent
these difficulties presenting a modal interpretation of quantum
logic in terms of its semantical analysis \cite{vF73}. In our case,
the modal component was introduced with different purposes: to
provide a rigorous framework for the Born rule and mainly, to
discuss the restrictions posed by the KS theorem to modalities
\cite{DFD1}.

To do so we enriched the orthomodular structure with a modal
operator taking into account the following considerations:

1) Propositions about the properties of the physical system are
interpreted in the orthomodular lattice of closed subspaces of the
Hilbert space of the (pure) states of the system. Thus we will
retain this structure in our extension.

2) Given a proposition about the system, it is possible to define a
context from which one can predicate with certainty about it
together with a set of propositions that are compatible with it and,
at the same time, predicate probabilities about the other ones (Born
rule). In other words, one may predicate truth or falsity of all
possibilities at the same time, i.e. possibilities allow an
interpretation in a Boolean algebra. In rigorous terms, for each
proposition $P$, if we refer with $\Diamond P$ to the possibility of
$P$, then $\Diamond P$ will be a central element of a orthomodular
structure.

3) If $P$ is a proposition about the system and $P$ occurs, then it
is trivially possible that $P$ occurs. This is expressed as $P \leq
\Diamond P$. \hspace{0.2cm}

4) Assuming an actual property and a complete set of properties that
are compatible with it determines a context in which the classical
discourse holds. Classical consequences that are compatible with it,
for example probability assignments to the actuality of other
propositions, shear the classical frame. These consequences are the
same ones as those which would be obtained by considering the
original actual property as a possible one. This is interpreted in
the following way: if $P$ is a property of the system, $\Diamond P$
is the smallest central element greater than $P$.

From consideration 1) it follows that the original orthomodular
structure is maintained. The other considerations are satisfied if
we consider  a modal operator $\Diamond$ over an orthomodular
lattice $L$ defined as $\Diamond a = Min \{z\in Z(L): a\leq z \}$
with $Z(L)$ the center of $L$ under the assumption that this minimum
exists for every $a\in L$. In the following section we explicitly
show our construction.  For technical reasons this algebraic study
will be performed using the necessity operator $\Box$  instead of
the possibility operator $\Diamond$. As usual, it will be then
possible to define the possibility operator from the necessity
operator.

\section{Orthomodular structures and modality }

\begin{definition}
{\rm Let $A$ be an orthomodular lattice. We say that $A$ is {\it
Boolean saturated} if and only if for all $a\in A$ the set $\{z\in
Z(A): z\leq a \}$ has a maximum. In this case such maximum is
denoted by $\Box (a)$. }
\end{definition}

\begin{example}
{\rm In view of Proposition \ref{eqcentro}, orthomodular complete
lattices considering $e(a) = \bigvee \{z \in Z(L) : z \leq a \}$ as
operator, are examples of boolean saturated orthomodular lattices.}
\end{example}

\begin{prop}\label{PROST}
Let $A$ be an orthomodular lattice. Then $A$ is boolean saturated
iff  there exists an unary operator $\Box$ satisfying

\begin{enumerate}

\item[S1]
$\Box x \leq x$

\item[S2]
$\Box 1 = 1$

\item[S3]
$\Box \Box x = \Box x$

\item[S4]
$\Box(x \land y) = \Box(x) \land \Box(y)$

\item[S5]
$y = (y\land \Box x) \lor (y \land \neg \Box x)$

\item[S6]
$\Box (x \lor \Box y ) = \Box x \lor \Box y $

\item[S7]
$\Box (\neg x \lor (y \land x)) \leq \neg \Box x \lor \Box y $

\end{enumerate}
\end{prop}

\begin{proof}
Suppose that $A$ is is Boolean saturated. S1), S2) and S3) are
trivial. \hspace{0.2cm} S4) Since $x\land y \leq x$ and $x\land y
\leq y$ then $\Box(x\land y) \leq \Box(x) \land \Box(y)$. For the
converse, $\Box(x) \leq x$ and $\Box(y) \leq y$, thus $\Box(x) \land
\Box(y) \leq \Box(x\land y)$. \hspace{0.2cm} S5) Follows from
Proposition \ref{eqcentro} since $\Box(x) \in Z(A)$. \hspace{0.2cm}
S6) For simplicity, let $z = \Box y$. From the precedent item and
taking into account that $z \in Z(L)$ we have that $\Box(z\lor x)
\land \Box(\neg z \lor x) = \Box ((z\lor x)\land (\neg z \lor x )) =
\Box(x)$. Since  $\neg z \leq \Box(\neg z \lor x)$ then we have that
$1 = z \lor \neg z \leq z \lor \Box(\neg z \lor x)$. Also we have
$z\leq \Box(z\lor x)$. Finally $z\lor \Box(x) = (z\lor \Box(z\lor
x)) \land (z\lor \Box(\neg z\lor x)) = (z\lor \Box(z\lor x)) \land 1
= \Box(z\lor x)$ i.e. $\Box (x \lor \Box y ) = \Box x \lor \Box y $.
\hspace{0.2cm} S7) Since $\Box(x) \leq x$ then $\neg x \leq \neg
\Box x$, we have that $\neg x \lor (y \land x) \leq \neg \Box x \lor
y$. Using the precedent item $\Box (\neg x \lor (y \land x))  \leq
\Box (\neg \Box x \lor y) = \neg \Box x \lor
\Box y$ since $\neg \Box x \in Z(A)$. \\

For the converse, let $a\in A$ and $\{z \in Z(A) : z\leq a \}$. By
$S1$ and $S5$ it is clear that $\Box a \in \{z \in Z(A) : z\leq a
\}$. We see that $\Box a$ is the upper bound of the set. Let $z\in
Z(A)$ such that $z \leq a$ then $1 = \neg z \lor (a \land z)$. Using
$S2$ and $S7$ we have $1 = \Box 1 = \Box (\neg z \lor (a \land z))
\leq \neg \Box z \lor a = \neg z \lor a $. Therefore $z = z \land
(\neg z \lor \Box a )$ and since $z$ is central $z = z \land \Box a$
resulting $z \leq \Box a$. Finally $ \Box a = Max\{z \in Z(A) :
z\leq a \} $. \qed
\end{proof}\\

Note that the operator $\Box$ is an example of quantifier in the
sense of Janowitz \cite{JAN}.

\begin{theo}
The class of Boolean saturated orthomodular lattices constitutes a
variety which is axiomatized by

\begin{enumerate}
\item
Axioms of ${\cal OML}$,

\item
$S1,...,S7$.

\end{enumerate}
\end{theo}

\begin{proof}
Obvious by Proposition \ref{PROST}
\qed \\
\end{proof}

Boolean saturated orthomodular lattices  are algebras $ \langle A,
\land, \lor, \neg, \Box, 0, 1  \rangle$ of type $ \langle 2, 2, 1,
1, 0, 0 \rangle$. The variety of this algebras is noted as  ${\cal
OML}^\Box$. By simplicity, the set $Term_{{\cal OML}^\Box}$ will be
denoted by $Term^\Box$. Since ${\cal OML}$ is a reduct of ${\cal
OML}^{\Box}$ we can also suume that all ${\cal
OML}^{\Box}$-equations are of the form $t=1$. It is well known that
${\cal OML}$ is congruence distributive and congruence permutable.
Therefore if $A \in {\cal OML}^{\Box}$ and we consider the
OML-reduct of $A$ it is clear that $Con_{{{\cal OML}^{\Box}}}(A)
\subseteq Con_{{\cal OML}}(A) $ resulting $A$ congruence
distributive and congruence permutable in the sence of ${\cal
OML}^{\Box}$-congruences. Hence the variety ${\cal OML}^{\Box}$ is
congruence distributive and congruence permutable. The following
lemma gives basic properties that will be used later:

\begin{lem} \label{PROPERTY}
Let $A \in {\cal OML}^{\Box} $ and $a, b \in A$ and $z_1, z_2 \in
Z(A)$. Then we have:

\begin{enumerate}

\item
$\neg \Box a \lor a = 1$,

\item
$\neg(a\lor \neg b) \lor (a \lor \neg \Box b) = 1$,

\item
$\neg (\neg z_1 \lor z_2) \lor ((\neg (z_1 \lor a) \lor (z_2 \lor a)
= 1$

\item
$\Box a \lor \Box b \leq \Box(a\lor b) $,

\item
$(\neg \Box a \land \neg \Box b) \lor \Box (a\lor b) = 1$,

\item
if $x\leq y$ then $\Box x \leq \Box y$.

\end{enumerate}
\end{lem}

\begin{proof}
1) \hspace{0.2cm} Since $\Box a \leq a$ then $\neg a \leq \neg \Box
a$ and $1 = a \lor \neg  a \leq a \lor \neg \Box a$. \hspace{0.2cm}
2) \hspace{0.2cm} Since $\neg \Box b \in Z(A)$ and by item 1 we have
that  $\neg(a\lor \neg b) \lor (a \lor \neg \Box b) = (\neg a \land
b) \lor (a \lor \neg \Box b) = ((\neg a \lor \neg \Box b) \land (b
\lor \neg \Box b)) \lor a = ((\neg a \lor \neg \Box b) \land 1) \lor
a  = 1$. \hspace{0.2cm} 3) \hspace{0.2cm} $\neg (\neg z_1 \lor z_2)
\lor ((\neg (z_1 \lor a) \lor (z_2 \lor a) = \neg ((\neg z_1 \lor
z_2) \land (z_1 \lor a))\lor (z_2 \lor a) = \neg( (\neg z_1 \land a)
\lor (z_1 \land z_2) \lor (z_2 \land a)) \lor (z_2 \lor a) = ((z_1
\lor \neg a) \land (\neg z_1 \lor \neg z_2) \land (\neg z_2 \lor
\neg a)) \lor (z_2 \lor a) = ((z_1 \lor \neg a \lor z_2) \land (\neg
z_2 \lor \neg x \lor z_2) \land (\neg z_2 \lor \neg a \lor z_2))
\lor a = z_1 \lor \neg a \lor z_2 \lor a = 1$.  \hspace{0.2cm} 4)
\hspace{0.2cm}$\Box a \leq a$ and $\Box b \leq b$, $\Box a  \lor
\Box b \leq a \lor b$. Since $\Box a  \lor \Box b \in Z(A)$ it is
clear that  $\Box a  \lor \Box b \leq \Box (a  \lor b)$.
\hspace{0.2cm} 5) \hspace{0.2cm} Immediately from item 4.
\hspace{0.2cm} 6) \hspace{0.2cm} Suppose that $x\leq y$ then $x =
x\land y$. By Axiom $S4$ we have that $\Box x = \Box (x\land y) =
\Box x\land  \Box y)$, hence $\Box x \leq \Box y$. \qed
\end{proof}

\begin{lem}\label{FACTOR1}
Let $A \in {\cal OML}^{\Box}$ and $z \in Z(A)$. Then the binary
relation ${\Theta}_z$ on $A$ defined by $a  \Theta_z b$  iff $a\land
z = b\land z$ is a congruence on $A$,   such that $A\cong
A/{\Theta}_z\times A/{\Theta}_{\neg z}$.
\end{lem}

\begin{proof}
It is well known that $\Theta_z$ is a ${\cal OML}$-congruence and
$A$ is ${\cal OML}$-isomorphic to $A/{\Theta}_z\times
A/{\Theta}_{\neg z}$. Therefore we only need to see the
$\Box$-compatibility. In fact: suppose that $a \Theta_z b$ then
$a\land z = b\land z$. Therefore $\Box(a) \land z = \Box(a) \land
\Box(z) = \Box (a\land z) = \Box (b\land z) = \Box(b) \land \Box(z)
= \Box(b) \land z$. Hence $\Box(a) \Theta_z \Box (b)$. \qed
\end{proof}

\begin{prop}\label{FACTOR2}
Let $A \in {\cal OML}^{\Box}$ then we have that:

\begin{enumerate}
\item
The map $z \rightarrow \Theta_z$ is a lattice isomorphism between
$Z(L)$ and the Boolean subalgebra of $Con(L)$ of factor congruences.

\item
$A$ is directly indecomposable iff $Z(A) = \{0,1\}$.

\end{enumerate}
\end{prop}

\begin{proof}
1) Follows from Lemma \ref{FACTOR1} using the same arguments that
prove the analog result for orthomodular lattices {\rm
\cite[Proposition 5.2]{BH}}. \hspace{0.2cm} 2) Follows form item 1.
\qed
\end{proof}

\begin{prop}\label{DISC1}
Let $A$ be a directly indecomposable ${\cal OML}^{\Box}$-algebra.
Then $$t(x,y,z) =  (x \land \neg \Box ((x\land y) \lor (\neg x \land
\neg y))) \lor  (z \land \Box ((x\land y) \lor (\neg x \land \neg
y)))$$ is a discriminator term for $A$.

\end{prop}

\begin{proof}
By Proposition \ref{FACTOR2}, $Z(A) = \{0,1\}$. Therefore for each
$a \in A-\{1\}$, $\Box(a) = 0$. Let $a,b,c \in A$. Suppose that
$a\not = b$. By the characterization of the equations in ${\cal
OML}^{\Box}$ we have that $(a\land b) \lor (\neg a \land \neg b)
\not = 1$ and then $t(a,b,c) = a$. If we suppose that $a=b$ then it
is clear that $t(a,b,c) = c$. Hence $t(x,y,z)$ is a discriminator
term for $A$. \qed
\end{proof}

\begin{prop}\label{DISC2}
If ${\cal A}$ is a subvariety of ${\cal OML}^{\Box}$ then ${\cal A}$
is a discriminator variety.

\end{prop}

\begin{proof}
Let $SI_{\cal A}$ be the class of subdirectly irreducible algebras
of ${\cal A}$. Each algebra of $SI_{\cal A}$ is directly
indecomposable. Therefore by  Proposition \ref{DISC1} $SI_{\cal A}$
admit a common discriminator term. Since ${\cal A} = {\cal
V}(SI_{\cal A})$ we have that ${\cal A}$ is a discriminator variety.
\qed
\end{proof}

\section {Hilbert-style calculus for ${\cal OML}^\Box$}

In this section we build a Hilbert-style calculus $\langle
Term^{\Box}, \vdash \rangle$ for ${\cal OML}^\Box$. We first
introduce some notation. $\alpha \in Term^{\Box}$ is a {\it
tautology} iff $\models_{{\cal OML}^\Box} \alpha = 1$. Each
subset $T$ of $Term^{\Box}$ is referred as {\it theory}. If $v$ is a
valuation, $v(T) = 1$ means that $v(\gamma)= 1$ for each $\gamma \in
T$.  We use $T \models_{{\cal OML}^\Box} \alpha$ (read $\alpha$ is
{\it semantic consequence} of $T$) in the case in which when $v(T) =
1$ then $v(\alpha) = 1$ for each valuation $v$.

\begin{lem} \label{DED1}
Let $\gamma$ and $\alpha \in Term^\Box$. Then we have

\begin{enumerate}
\item
If $v$ is a valuation then $v(\alpha) = 1$ iff $v(\Box \alpha) = 1$.

\item
$\gamma \models_{{\cal OML}^\Box} \alpha$  iff \hspace{0.02cm}
$\gamma \models_{{\cal OML}^\Box} \Box \alpha$ \hspace{0.02cm} iff
\hspace{0.02cm} $\Box \gamma \models_{{\cal OML}^\Box} \alpha$
\hspace{0.02cm} iff \hspace{0.02cm} $\Box \gamma \models_{{\cal
OML}^\Box} \Box \alpha$.

\end{enumerate}
\end{lem}

\begin{proof}
1) If $v(\alpha) = 1$ then $1 = \Box 1 = \Box(v(\alpha)) = v(\Box
\alpha)$. The converse follows from the fact $1 = v(\Box \alpha) =
\Box(v(\alpha)) \leq v(\alpha)$. \hspace{0.2cm} 2) Immediate from
the item 1. \qed
\end{proof}

\begin{definition}
{\rm Consider by  definition the following binary connective
$$\alpha R \beta \hspace{0.2 cm} \mbox{for} \hspace{0.2 cm} (\alpha \land \beta) \lor (\neg \alpha\land \neg \beta)$$
The calculus $\langle Term^{\Box}, \vdash \rangle$ is given by the
following axioms:

\begin{enumerate}
\item[A0]
$1R(\alpha \lor \neg \alpha)$ and $0 R (\alpha \land \neg \alpha) $,

\item[A1]
$\alpha R \alpha$,

\item[A2]
$\neg(\alpha R \beta) \lor (\neg(\beta R \gamma)\lor (\alpha R
\gamma))$,

\item[A3]
$\neg(\alpha R \beta) \lor (\neg \alpha R \neg \beta) $,

\item[A4]
$\neg(\alpha R \beta) \lor ((\alpha \land \gamma) R (\beta \land
\gamma))$,

\item[A5]
$(\alpha \land \beta) R (\beta \land \alpha)$,

\item[A6]
$(\alpha \land (\beta \land \gamma)) R ((\alpha \land \beta) \land
\gamma)$,

\item[A7]
$(\alpha \land (\alpha \lor \beta)) R \alpha$,

\item[A8]
$(\neg \alpha \land \alpha) R ((\neg \alpha \land \alpha)\land
\beta)$,

\item[A9]
$\alpha R \neg \neg \alpha$,

\item[A10]
$\neg(\alpha \lor \beta)R(\neg \alpha \land \neg \beta)$,

\item[A11]
$(\alpha \lor (\neg \alpha \land (\alpha \lor \beta)) R (\alpha \lor
\beta)$,

\item[A12]
$(\alpha R \beta) R (\beta R \alpha)$,

\item[A13]
$\neg(\alpha R \beta) \lor (\neg\alpha \lor \beta)$,

\item[A14]
$(\Box \alpha \lor \alpha) R \alpha$,

\item[A15]
$\Box(\alpha \lor \neg \alpha) R  (\alpha \lor \neg \alpha)$,

\item[A16]
$\Box \Box \alpha R \Box\alpha $,

\item[A17]
$\Box(\alpha \land \beta) R (\Box \alpha \land \Box \beta )$,

\item[A18]
$((\alpha \land \Box \beta)\lor (\alpha \land \neg \Box \beta)) R
\alpha $,

\item[A19]
$\Box(\alpha \lor \neg \Box \beta) R (\Box \alpha \lor \neg \Box
\beta)$,

\item[A20]
$\Box(\alpha \lor \Box \beta) R (\Box \alpha \lor \Box \beta)$,

\item[A21]
$(\Box(\neg \alpha \lor (\beta \land \alpha )) \lor (\neg \Box
\alpha \lor \Box \beta)) R (\neg \Box \alpha \lor \Box \beta) $,

\item[A22]
$\neg(\alpha \lor \neg \beta) \lor (\alpha \lor \neg \Box \beta)$,

\item[A23]
$\neg (\gamma \lor \neg \beta) \lor ( \neg (\beta \lor \alpha) \lor
(\gamma \lor \alpha))$,

\item[A24]
$\Box(\alpha \lor \beta)\lor (\neg \Box \alpha \land  \neg \Box
\beta )$.

\end{enumerate}

\noindent and the following inference rules:

$$
\displaylines{ \hfill {\alpha, \neg \alpha \lor \beta \over \beta}
\hfill \llap{\it disjunctive syllogism (DS)} \cr\cr \hfill {\alpha
\over \Box \alpha}, \hfill \llap{\it necessitation (N)} }
$$
}
\end{definition}

Let $T$ be a theory.  A {\it proof}  from $T$ is a sequence
$\alpha_1,...,\alpha_n$ in $Term^{\Box}$ such that each member is
either an axiom or a member of $T$ or follows from some preceding
member of the sequence using {\it DS} or  {\it N}. $T \vdash \alpha$
means that  $\alpha$ is provable in $T$, that is, $\alpha$ is the
last element of a proof from $T$. If $T = \emptyset$, we use the
notation $\vdash \alpha$ and in this case we will say that $\alpha$
is a theorem of $\langle Term^{\Box}, \vdash \rangle$. $T$ is {\it
inconsistent} if and only if $T\vdash \alpha$ for each $\alpha \in
Term^{\Box}$; otherwise it is {\it consistent}.

\begin{prop}\label{COR}
Let $T$ be a theory and $\alpha, \beta, \gamma \in Term^{\Box}$.
Then we have

\begin{enumerate}
\item
$T \vdash \alpha R \beta \Longrightarrow  T\vdash \beta  R \alpha $

\item
$T \vdash \alpha R \beta \hspace{0.2cm} and \hspace{0.2cm} T \vdash
\beta R \gamma  \hspace{0.2cm} \Longrightarrow \hspace{0.2cm} T
\vdash \alpha  R \gamma $

\item
$T \vdash \alpha R \beta \hspace{0.2cm} \Longrightarrow
\hspace{0.2cm} T \vdash \neg \alpha R \neg \beta$

\item
$T \vdash \alpha R \beta \hspace{0.2cm} and \hspace{0.2cm} T \vdash
\alpha \land \gamma  \hspace{0.2cm} \Longrightarrow \hspace{0.2cm} T
\vdash \beta  \land \gamma $

\item
$T \vdash \alpha R \beta \hspace{0.2cm} and \hspace{0.2cm} T \vdash
\alpha \lor \gamma  \hspace{0.2cm} \Longrightarrow \hspace{0.2cm} T
\vdash \beta  \lor \gamma $

\item
$T \vdash \alpha R \beta \hspace{0.2cm} \Longrightarrow
\hspace{0.2cm} T \vdash \Box \alpha R \Box \beta$

\item
$ \vdash \alpha \lor \neg \alpha$

\item
$T \vdash \alpha \hspace{0.2cm} \Longrightarrow \hspace{0.2cm} T
\vdash \alpha  \lor \beta$

\end{enumerate}
\end{prop}

\begin{proof}

\noindent 1)
\begin{enumerate}
\item[(1)]
$T \vdash \alpha R \beta$

\item[(2)]
$T \vdash (\alpha R \beta) R (\beta R \alpha)$ \hspace{5.9 cm}
{\footnotesize by A12}

\item[(3)]
$T \vdash \neg ((\alpha R \beta) R (\beta R \alpha)) \lor (\neg
(\alpha R \beta) \lor (\beta R \alpha)) $ \hspace{1.8 cm}
{\footnotesize by A13}

\item[(4)]
$T \vdash (\neg (\alpha R \beta) \lor (\beta R \alpha)) $
\hspace{5.2 cm} {\footnotesize by $DS$ 2,2}

\item[(5)]
$T \vdash \beta R \alpha $ \hspace{7.6 cm} {\footnotesize by $DS$
1,4}

\end{enumerate}

\noindent 2) Is easily from A2 and two application of the ${\it
DS}$. \hspace{0.2 cm}

\noindent 3) Follows from A3. \hspace{0.2 cm}

\noindent 4)
\begin{enumerate}
\item[(1)]
$T \vdash \alpha R \beta$

\item[(2)]
$T \vdash \alpha \land \gamma$

\item[(3)]
$T \vdash \neg (\alpha R \beta) \lor ((\alpha \land \gamma)R(\beta
\land \gamma))  $ \hspace{3.7 cm} {\footnotesize by A4}

\item[(4)]
$T \vdash (\alpha \land \gamma)R(\beta \land \gamma)$ \hspace{5.8
cm} {\footnotesize by $DS$ 1,2}

\item[(5)]
$\neg (\alpha \land \gamma)R(\beta \land \gamma) \lor (\neg (\alpha
\land \gamma) \lor (\beta \land \gamma))$ \hspace{2.5 cm}
{\footnotesize by A4}

\item[(6)]
$T \vdash \beta \land \gamma$ \hspace{7.6 cm} {\footnotesize by $DS$
5,4,2}

\end{enumerate}

\noindent 5) Follows by item 4, A9 and A10. \hspace{0.2 cm}

\noindent 6)
\begin{enumerate}
\item[(1)]
$T \vdash \alpha R \beta$

\item[(2)]
$T \vdash (\alpha \land \beta) \lor (\neg \alpha \land \neg \beta)$
\hspace{4.7 cm} {\footnotesize equiv 1}

\item[(3)]
$T \vdash (\alpha \land \beta) \lor \neg (\alpha \lor \beta)$
\hspace{5 cm} {\footnotesize by item 5 and A10}

\item[(4)]
$\vdash \neg ((\alpha \land \beta) \lor \neg (\alpha \lor \beta))
\lor ((\alpha \land \beta) \lor \neg \Box (\alpha \lor \beta))$
\hspace{0.7 cm} {\footnotesize by A22}

\item[(5)]
$T \vdash (\alpha \land \beta) \lor \neg \Box (\alpha \lor \beta)$
\hspace{3.5 cm} {\footnotesize by $DS$ 4,3}

\item[(6)]
$T \vdash \Box ((\alpha \land \beta) \lor \neg \Box (\alpha \lor
\beta))$ \hspace{2.9 cm} {\footnotesize by $DS$ 4,3}

\item[(7)]
$T \vdash \Box (\alpha \land \beta) \lor \neg \Box (\alpha \lor
\beta)$ \hspace{3.2 cm} {\footnotesize by A13, A19}

\item[(8)]
$T \vdash (\Box \alpha \land  \Box \beta) \lor \neg \Box (\alpha
\lor \beta)$ \hspace{2.9 cm} {\footnotesize by item 5 and A17}

\item[(9)]
$\neg ((\Box \alpha \land  \Box \beta) \lor \neg \Box (\alpha \lor
\beta)) \lor ( \neg (\Box (\alpha \lor \beta) \lor (\neg \Box \alpha
\land \neg \Box \beta)) \lor ( (\Box \alpha \land  \Box \beta) \lor
(\neg \Box \alpha \land \neg \Box \beta))$ \hspace{4.1 cm}
{\footnotesize by  A23}

\item[(10)]
$\Box(\alpha \lor \beta)\lor (\neg \Box \alpha \land  \neg \Box
\beta )$ \hspace{3.5 cm} {\footnotesize by A24}

\item[(11)]
$(\Box \alpha \land  \Box \beta) \lor (\neg \Box \alpha \lor \neg
\Box \beta))$ \hspace{3 cm} {\footnotesize by $SD$ 8,9,10}

\item[(12)]
$T \vdash \Box \alpha R \Box \beta$ \hspace{5.6 cm} {\footnotesize
equiv 1}

\end{enumerate}

\noindent
7)Follows from A1 and A13. \\
\noindent 8)
\begin{enumerate}
\item[(1)]
$T \vdash \alpha $

\item[(2)]
$T \vdash (\alpha \lor \neg \alpha) R ((\alpha \lor \neg \alpha)
\lor \beta) $ \hspace{3.1 cm} {\footnotesize by A3, A8, A10 }

\item[(3)]
$T \vdash (\alpha \lor \neg \alpha) \lor \beta $ \hspace{5.2 cm}
{\footnotesize by item 7 and A13}

\item[(4)]
$\vdash \neg ((\alpha \land \beta) \lor \neg (\alpha \lor \beta))
\lor ((\alpha \land \beta) \lor \neg \Box (\alpha \lor \beta))$
\hspace{0 cm} {\footnotesize by A22}

\item[(5)]
$\vdash \neg \alpha \lor (\alpha \lor \beta) $ \hspace{5.6 cm}
{\footnotesize by 4, A5, A3, A10}

\item[(6)]
$\vdash \alpha \lor \beta $ \hspace{6.9 cm} {\footnotesize by $DS$
1, 5 } \qed

\end{enumerate}

\end{proof}

\begin{prop} \label{TAUT}
Axioms of the $\langle Term^{\Box}, \vdash \rangle$ are tautologies.

\end{prop}

\begin{proof}
For A0... A13 see {\rm (\cite[Chapter 4.15 ]{KAL})}. A22...24 follow
from Proposition \ref{PROPERTY}. \qed
\end{proof}

\begin{theo}\label{LINDENBAUM}
Let $T$ be a theory. If for each $\alpha \in Term^{\Box} $ we
consider the set $[\alpha] = \{\beta : T\vdash \alpha R \beta \}$
then $L_T = \{[\alpha] : \alpha \in Term^{\Box} \}$ determines a
partition in equivalence classes of $Term^{\Box}$. Defining the
following operation in $L_T$

\begin{enumerate}

\item[]
$[\alpha]\land [\beta] = [\alpha \land \beta] $  \hspace{0.7cm}
$\neg[\alpha] = [\neg \alpha]$  \hspace{0.7cm}  $0 = [0]$

\item[]
$[\alpha]\lor [\beta] = [\alpha \lor \beta] $   \hspace{0.7cm}
$\Box[\alpha] = [\Box \alpha]$   \hspace{0.6cm} $1 = [1]$

\end{enumerate}

\noindent then we have

\begin{enumerate}
\item
$ \langle L_T, \lor, \land, \neg, \Box, 0, 1 \rangle$ is a Boolean
saturated orthomodular lattice.

\item
$T\vdash \alpha$ if and only if $[\alpha] = 1$

\end{enumerate}

\end{theo}

\begin{proof}
1) By A1 and Proposition \ref{COR} (item 1 and 2)  $L_T = \{[\alpha]
: \alpha \in Term^{\Box} \}$ is a partition in equivalence classes
of $Term^{\Box}$. By Proposition \ref{COR} (item 3 and 6) $\lor,
\land, \neg, \Box$ are well defined in $L_T$. By A0...A13 and  {\rm
(\cite[Proposition 4.15. 1]{KAL})} $L_T$ is an orthomodular lattice.
By A14...A21 and Proposition \ref{TAUT}, $L_T$ is boolean saturated.
\hspace{0.2cm} 2) Assume that $T\vdash \alpha$, then we have that:

\begin{enumerate}

\item[(1)]
$T \vdash \alpha $

\item[(2)]
$T \vdash \alpha R (\alpha \land (\alpha \lor \neg \alpha)) $
\hspace{3.3 cm} {\footnotesize by A7}

\item[(3)]
$T \vdash \alpha \land (\alpha \lor \neg \alpha)  $ \hspace{4.2 cm}
{\footnotesize by 1 and A13}

\item[(4)]
$T \vdash (\alpha \land (\alpha \lor \neg \alpha)) \lor (\neg \alpha
\land  \neg (\alpha \lor \neg \alpha))   $ \hspace{0.5 cm}
{\footnotesize by 3 and Prop. \ref{COR} 8}

\item[(5)]
$T \vdash \alpha R (\alpha \lor \neg \alpha) $ \hspace{4.3 cm}
{\footnotesize equiv in 4}

\end{enumerate}

\noindent resulting $[\alpha] = 1$. On the other hand, if $[\alpha]
= 1$, we have that $T \vdash \alpha R (\alpha \lor \neg \alpha) $.
Using Proposition \ref{COR} 7 and Ax13, it results $T\vdash \alpha$.
\qed
\end{proof}\\

\noindent The following theorem establishes the strong completeness
for $\langle Term^{\Box}, \vdash \rangle$ with respect to the
variety ${\cal OML}^\Box$.

\begin{theo} \label{COM}
Let $\alpha \in Term^{\Box}$ and $T$ be a theory. Then we have that:
$$T \vdash \alpha \Longleftrightarrow T\models_{{\cal OML}^\Box} \alpha   $$
\end{theo}

\begin{proof}
If $T$ is inconsistent, this result is trivial. Assume that  $T$ is
consistent. $\Longrightarrow$) Immediate. \hspace{0.2cm}
$\Longleftarrow$) Suppose that $T$ does not prove $\alpha$. Then, by
Proposition \ref{LINDENBAUM}, $[\alpha] \neq 1$. Then the projection
$p:Term^\Box \rightarrow L_T $ with $p(\varphi) = [\varphi]$ is a
valuation such that $p(\varphi) = 1$ for each $\varphi \in T$ and
$p(\alpha) \neq 1$. Finally we have that not $T\models_{{\cal
OML}^\Box} \alpha$. \qed
\end{proof}

\begin{coro} \label{COMPACT} {\rm (Compactness)}
Let $\alpha \in Term^{\Box}$ and $T$ be a theory. Then we have that,
$T\models_{{\cal OML}^\Box} \alpha$ iff there exists a finite subset
$T_0 \subseteq T$
 such that $T_0 \models_{{\cal OML}^\Box} \alpha$.
\end{coro}

\begin{proof}
In view of Theorem \ref{COM}, if $T\models_{{\cal OML}^\Box} \alpha$
then $T\vdash \alpha$. If $\varphi_1, \cdots \varphi_m, \alpha$ is a
proof of $ \alpha$ from $T$, we can consider the finite set $T_0 =
\{ \varphi_i \in T : \varphi_i \in \{\alpha_1, \cdots \alpha_n\}\}$.
Using again Theorem \ref{COM} we have $T_0 \models_{{\cal OML}^\Box}
\alpha $. \qed
\end{proof}\\

We can also establish a kind of deduction theorem.

\begin{coro} \label{DED2}
Let $\gamma, \alpha \in Term^{\Box}$ and $T$ be a theory. Then we
have that: $$T \cup \{\gamma \} \vdash \alpha \hspace{0.3cm} iff
\hspace{0.3cm} T \vdash \neg \Box \gamma \lor \alpha$$
\end{coro}

\begin{proof}
By Theorem \ref{COM} we will prove that $T \cup \{\gamma \}
\models_{{\cal OML}^\Box} \alpha$ iff $T \models_{{\cal OML}^\Box}
\neg \Box \gamma \lor \alpha$. By Corollary \ref{COMPACT} $T \cup
\{\gamma \} \models_{{\cal OML}^\Box} \alpha$ iff there exists
$\varphi_1 \ldots \varphi_n \in T$ such that $(\varphi_1 \land
\ldots \land \varphi_n) \land \gamma \models_{{\cal OML}^\Box}
\alpha$. Let $\varphi = \varphi_1 \land \ldots \land \varphi_n$.
Then $\varphi \land \gamma \models_{{\cal OML}^\Box} \alpha$ implies
that $(\varphi \land \gamma) \lor \neg \Box(\gamma) \models_{{\cal
OML}^\Box} \neg \Box \gamma \lor \alpha$ and then $\varphi \lor \neg
\Box\gamma  \models_{{\cal OML}^\Box} \neg \Box\gamma \lor \alpha$
since for each valuation $v$, $v(\Box\gamma)$ is a central element
and $v(\gamma \lor \neg \Box\gamma)=1$. Since $\varphi
\models_{{\cal OML}^\Box} \varphi \lor \neg \Box\gamma$ we have that
$\varphi \models_{{\cal OML}^\Box} \Box\gamma \lor \alpha$ thus $T
\models_{{\cal OML}^\Box} \neg \Box\gamma \lor \alpha$.

On the other hand, if $T \models_{{\cal OML}^\Box} \neg \Box \gamma
\lor \alpha$ we can consider again $\varphi = \varphi_1 \land \ldots
\land \varphi_n$ such that $\varphi_1 \ldots \varphi_n \in T$ and
$\varphi \models_{{\cal OML}^\Box} \neg \Box \gamma \lor \alpha$.
Therefore $\varphi \land \Box \gamma \models_{{\cal OML}^\Box} \Box
\gamma \land (\neg \Box \gamma \lor \alpha)$ and then $\varphi \land
\Box \gamma \models_{{\cal OML}^\Box} \Box \gamma \land \alpha $
since for each valuation $v$, $v(\Box \gamma \land (\neg \Box \gamma
\lor \alpha)) = v(\Box \gamma \land \alpha)$ taking into account
that $v(\Box \alpha)$ is always a central element. Since $\Box
\gamma \land \alpha \models_{{\cal OML}^\Box} \alpha$ we have that
$\varphi \land \Box \gamma \models_{{\cal OML}^\Box} \alpha$.
Applying Lemma \ref{DED1} we have that  $\Box (\varphi \land \Box
\gamma) \models_{{\cal OML}^\Box} \alpha$ hence $\Box \varphi \land
\Box \gamma \models_{{\cal OML}^\Box} \alpha$ and $\varphi \land
\gamma \models_{{\cal OML}^\Box} \alpha$ in view of Axiom $S4$ of
${\cal OML}^\Box$. Thus $T \cup \{\gamma \} \models_{{\cal
OML}^\Box} \alpha$. \qed
\end{proof}

\section{Modal orthomodular frames and Kripke-style semantics}

In order to establish a Kripke-style semantics $\langle Term^{\Box},
\vdash \rangle$ we first introduce el concept of modal Baer
semigroups which constitute a sub-class of Baer $^\star$-semigroups.

\begin{definition}
{\rm A {\it modal Baer semigroup} is a Baer $^\star$-semigroup $G$
such that $\langle P_c(G), \land, \lor, ', 0,1  \rangle$ is a
Boolean saturated orthomodular lattice. A {\it modal orthomodular
frame}  is a pair $\langle G, u \rangle$ such that $G$ is a modal
Baer semigroup and $u$ is a valuation $u: Term^{\Box} \rightarrow
P_c(G)$ }
\end{definition}

We denote by ${\cal MOF}$ the class of all modal orthomodular
frames. The following result is a representation theorem by modal
Baer semigroups of Boolean saturated orthomodular lattices.

\begin{theo} \label{KRIPALG2}
Let $A \in {\cal OML}^\Box$, then there exists a modal Baer
semigroup $G(A)$ such that $A$ is ${\cal OML}^\Box$-isomorphic to
$P_c(G(A))$.
\end{theo}

\begin{proof}
Let $A \in {\cal OML}^\Box$. By Theorem \ref{PRO2} there exists a
Baer $^\star$-semigroup $G$ such that  $A$ is ${\cal
OML}$-isomorphic to $P_c(G(A))$. Since ${\cal OML}$-isomorphisms
preserve supremum of central elements we have that $P_c(G(A)) \in
{\cal OML}^\Box$ and then, $G(A) \in {\cal MBS}$. \qed
\end{proof}\\

Note that we can easily prove that $\models_{{\cal OML}^\Box} t
= 1$ iff for all modal Baer semigroups $G$ we have that
$\models_{P_c(G)} t = 1$.

\begin{prop} \label{KRIPALG2}
Let $\langle G, u \rangle$ be a modal orhomodular frame and $t,s \in
Term^\Box$. Then we have that:

\begin{enumerate}

\item
$u(t \land s)\cdot G = (u(t)\cdot G) \cap (v(t)\cdot G)$,

\item
$u(\neg t)\cdot G = \{x\in G: \forall y \in u(t) \cdot G,
\hspace{0.2cm} y^\star \cdot x = 0 \}$

\item

$u(\Box t) \cdot G  = \bigcup \{x\cdot G: x\in
Z(P_c(G))\hspace{0.2cm} and \hspace{0.2cm} x \leq v(t) \}$.

\end{enumerate}
\end{prop}

\begin{proof}
1) Follows from an analogous argument used in {\rm \cite[Theorem
3.13]{YM}}. \hspace{0.2cm} 2) See the proof of {\rm \cite[Lemma
3.16-3]{YM}}. \hspace{0.2cm} 3)We first note that $u(t) \geq \Box
u(t) =  u(\Box t) \in Z(P_c(G)) $. Thus $u(\Box t)\cdot G \in
\{x\cdot G: x\in Z(P_c(G))\hspace{0.2cm} and \hspace{0.2cm} x \leq
u(t) \}$ and then $u(t)\cdot G \subseteq \bigcup \{x\cdot G: x\in
Z(P_c(G))\hspace{0.2cm} and \hspace{0.2cm} x \leq u(t) \}$. On the
other hand, if $x\in Z(P_c(G))$ and $x \leq u(t)$ then $x\cdot G
\subseteq u(\Box t)\cdot G$ since $x\leq u(\Box t)$. Hence $\bigcup
\{x\cdot G: x\in Z(P_c(G))\hspace{0.2cm} and \hspace{0.2cm} x \leq
u(t) \} \subseteq u(t)\cdot G $. \qed
\end{proof}

\begin{definition}
{\rm Let $\langle G, u \rangle$ be a modal orhomodular frame. Then
we define inductively the {\it forcing relation}  $\models_{\langle
G, u \rangle}^x \subseteq G \times Term^{\Box}$ as follows:

\begin{enumerate}
\item
$\models_{\langle G, u \rangle}^x p$ \hspace{0.9cm} iff
\hspace{0.2cm} $x\in u(p)\cdot G$, for each variable $ p\in
Term^{\Box}$,

\item
$\models_{\langle G, u \rangle}^x \alpha \land \beta$ \hspace{0.2cm}
iff \hspace{0.2cm} $\models_{\langle G, u \rangle}^x \alpha$ and
$\models_{\langle G, u \rangle}^x \beta$,

\item
$\models_{\langle G, u \rangle}^x \neg \alpha $ \hspace{0.6cm} iff
\hspace{0.2cm} $\forall g \in G $, $\models_{\langle G, u \rangle}^g
\alpha \Longrightarrow g'\cdot x = 0$,

\item
$\models_{\langle G, u \rangle}^x \Box \alpha $ \hspace{0.6cm} iff
\hspace{0.2cm} $x = z\cdot g$ such that $z\in Z(P_c(G))$ and
$\models_{\langle G, u \rangle}^z \alpha$.

\end{enumerate}
}
\end{definition}

The relation $\models_{\langle G, u \rangle}^x \alpha$ is read as
{\it $\alpha$ is true at the point $x$ in the {\it modal
orthomodular frame} $\langle G, u \rangle$} and by $\models_{\langle
G, u \rangle} \alpha$ we understand that for each $x\in G$,
$\models_{\langle G, u \rangle}^x \alpha$. Generalizing, if $T$ is a
theory, $\models_{\langle G, u \rangle} T$ means that, for each
$\beta \in T$ we have that $\models_{\langle G, u \rangle} \beta$.
With these elements we can establish a notion of {\it consequence}
in the Kripke-style sense that will be noted by $T \models_{{\cal
MOF}} \alpha$.
$$T \models_{{\cal MOF}} \alpha \hspace{0.3cm} iff \hspace{0.3cm}
\forall \langle G, u \rangle \in {\cal MOF}, \hspace{0.2cm}
\models_{\langle G, u \rangle} T \Longrightarrow \models_{\langle G,
u \rangle} \alpha$$

\noindent Let $\alpha \in Term^{\Box}$, $T$ be a theory and $\langle
G, u \rangle$ be an orthomodular frame. Then we consider the
following sets: $$\mid \alpha   \mid_{\langle G, u \rangle} = \{x
\in G: \models_{\langle G, u \rangle}^x \alpha \} $$ $$\mid T
\mid_{\langle G, u \rangle} = \bigcap_{\beta \in T} \mid  \beta
\mid_{\langle G, u \rangle } $$

\begin{prop} \label{KRIPALG3}
Let $\alpha \in Term^{\Box}$, $T$ be a theory and  $\langle G, u
\rangle$ be  a modal orthomodular frame. Then we have that:

\begin{enumerate}
\item
$\mid \alpha \mid_{\langle G, u \rangle} = u(\alpha)\cdot G$,

\item
$\models_{\langle G, u \rangle} T$ \hspace{0.2 cm} iff \hspace{0.2
cm} $\mid T\mid_{\langle G, u \rangle} = G$

\end{enumerate}
\end{prop}

\begin{proof}
1) We use induction on the complexity of terms. If $\alpha$ is a
variable the proposition results trivial. If $\alpha$ is $\beta
\land \gamma$ or $\neg \beta$ we refer to {\rm \cite[Lemma
3.16]{YM}}. Suppose that $\alpha$ is $\Box\beta$. We prove that
$u(\Box \beta)\cdot G \subseteq \hspace{0.2cm} \mid \Box \beta
\mid_{\langle G, u \rangle}$. By Proposition \ref{PROPG}-1 and by
inductive hypothesis, we have that $u(\Box \beta)\cdot G =
\Box(u(\beta)) \cdot G \subseteq u(\beta)\cdot G = \mid \beta \mid
_{\langle G, u \rangle}$. Then $\Box(u(\beta)) \cdot 1 =
\Box(u(\beta)) \in \mid \beta \mid _{\langle G, u \rangle}$ i.e.,
$\models_{\langle G, u \rangle}^{\Box(u(\beta))} \beta$. Thus if $x
\in u(\Box \beta)\cdot G$ then $x = \Box(u(\beta))\cdot g$ and,
taking into account that $\Box(u(\beta)) \in Z(P_c(G))$, it results
that $x\in \mid \Box \beta \mid_{\langle G, u \rangle}$. On the
other hand, if $x\in \hspace{0.2cm} \mid \Box \beta \mid_{\langle G,
u \rangle}$, then $x = z\cdot g$ such that $z\in Z(P_c(G))$ and
$z\in \hspace{0.2cm} \mid \beta \mid_{\langle G, u \rangle}$. Then
by inductive hypothesis we have that $z = u(\beta)\cdot g_0 \leq
u(\beta) \cdot 1 = u(\beta)$.
By the lattice-order definition of $\Box (u(\beta))$ it is clear that $z \leq \Box(u(\beta)) = u(\Box(\beta))$. Therefore $z\cdot G \subseteq u(\Box(\beta))\cdot G$. Hence $x \in u(\Box(\beta))\cdot G$. \\

2) Using the above item, $\models_{\langle G, u \rangle} T$ iff
$\forall \beta \in T$, $\models_{\langle G, u \rangle} \beta$ iff
$\forall \beta \in T$, $\mid \beta \mid_{\langle G, u \rangle} = G$
iff $G =  \bigcap_{\beta \in T} \mid  \beta \mid_{\langle G, u
\rangle } = \mid T \mid_{\langle G, u \rangle} $ \qed
\end{proof}

\begin{theo}\label{COM1}{\rm [Kripke style completeness]}
Let $\alpha \in Term^\Box $ and $T$ be a theory.  Then we have $$T
\models_{{\cal OML}^\Box} \alpha  \Longleftrightarrow T
\models_{{\cal MOF}} \alpha $$

\end{theo}

\begin{proof}
Suppose that $T\models_{{\cal OML}^\Box}\alpha$. By Corollary
\ref{COMPACT} there exists $\gamma_1, \ldots \gamma_n \in T$ such
that if we consider $\gamma$ as $\gamma_1 \land \ldots \land
\gamma_n $ then $\gamma \models_{{\cal OML}^\Box} \alpha$. By
Corollary \ref{DED2} we have that $\models_{{\cal OML}^\Box} \neg
\Box \gamma \lor \alpha$. Let $\langle G, u \rangle$ be  a modal
orthomodular frame such that  $\models_{\langle G, u \rangle} T$. By
Proposition \ref{KRIPALG3} and Proposition \ref{PROPG}-1 we have
that  $\mid \gamma \mid_{\langle G, u \rangle} = G$ and then
$u(\gamma) = 1$. But $u(\gamma) = 1$ implies $u(\Box\gamma) = 1$ and
$\neg u(\Box\gamma) = 0$. Therefore, necessarily $u(\alpha) = 1$,
and  $\mid \alpha \mid_{\langle G, u \rangle} = G$. Hence
$\models_{\langle G, u \rangle} \alpha$.

On the other hand we assume that $T \models_{{\cal MOF}} \alpha$.
Suppose that $T \not\models_{{\cal OML}^\Box} \alpha$. Then there
exists $A\in {\cal OML}^\Box$ and a valuation $v: Term^\Box
\rightarrow A$ such that $v(T) = 1$ but $v(\alpha) \not = 1$. By
Theorem \ref{KRIPALG2} there exists a modal Baer semigroup $G(A)$
such that $A$ is ${\cal OML}^\Box$-isomorphic to $P_c(G(A))$ being
$f: A \rightarrow P_c(G(A))$ such isomorphism. Consider the modal
orthomodular frame ${\langle G(A), fv \rangle}$. Then for each
$\beta \in T$ we have that $\mid \beta \mid_{{\langle G(A), fv
\rangle}} = fv(\beta)\cdot G(A) = 1\cdot G(A) = G(A)$. Therefore
$\mid T \mid_{{\langle G(A), fv \rangle}} = G(A)$ and
$\models_{\langle G(A), fv \rangle} T$ in view of Proposition
\ref{KRIPALG3}. By Proposition \ref{PROPG}-1  $\mid \alpha
\mid_{{\langle G(A), fv \rangle}} = fv(\alpha)\cdot G(A) \not =
G(A)$ again since $fv(\alpha) < 1$. Then $\not \models_{{\langle
G(A), fv \rangle}} \alpha$ which is a contradiction. Hence $T
\models_{{\cal OML}^\Box} \alpha$. \qed
\end{proof}

\section*{Conclusions}

We have developed a logical system based on the orthomodular
structure of propositions about quantum systems enriched with a
modal operator. We have obtained algebraic completeness and
completeness with respect to a Kripke-style semantic founded on Baer
$^\star$-semigroups. The importance of this structure from a
physical perspective deals with the interpretation of quantum
mechanics in terms of modalities.

{\small \noindent Graciela Domenech \\Instituto de Astronom\'{\i}a y
F\'{\i}sica del Espacio (IAFE-CONICET)\\CC 67 - Suc 28 -  1428
Buenos Aires -
Argentina\\e-mail: domenech@iafe.uba.ar}\\

{\small \noindent Hector Freytes \\Instituto Argentino de Matem\'{a}tica (IAM-CONICET) \\
Saavedra 15, 3er P - 1083 Buenos Aires - Argentina\\e-mail: hfreytes@gmail.com}\\

{\small \noindent Christian de Ronde \\Center Leo Apostel (CLEA) and Foundations of the Exact Sciences (FUND)\\
Brussels Free University - Krijgskudestraat 33, 1160 \\
Brussels-Belgium\\e-mail: cderonde@vub.ac.be}

\end{document}